\pdfoutput=1
\documentclass[prd,twocolumn,superscriptaddress,
preprintnumbers,nofootinbib]{revtex4}
\usepackage{graphicx}
\usepackage{epsfig}
\usepackage{bm}
\usepackage{latexsym,amssymb,amsmath,amsfonts,amssymb,txfonts,wasysym,float}
\usepackage{scrextend}
\usepackage{mathrsfs}
\usepackage{color}

\usepackage{enumitem}

\newcommand{\postscript}[2]{\setlength{\epsfxsize}{#2\hsize}
   \centerline{\epsfbox{#1}}}

\usepackage[usenames,dvipsnames]{xcolor}
\definecolor{orange}{cmyk}{0,0.5,1,0}
\definecolor{rossoCP3}{cmyk}{0,.88,.77,.40}
\definecolor{graa}{rgb}{0.8,0.8,0.8}
\definecolor{blaa}{rgb}{0.2,0.2,0.6}

\begin{document}

\title{\color{rossoCP3}  Necessary Conditions for Earthly Life Floating in the Venusian
Atmosphere}

\author{Jennifer J. Abreu}

\affiliation{Department of Physics and Astronomy,  Lehman College, City University of
  New York, NY 10468, USA
}

\author{Alyxander R. Anchordoqui}
\affiliation{John F. Kennedy School, Somerville, MA 02144, USA}

\author{Nyamekye J. Fosu}

\affiliation{Department of Physics and Astronomy,  Lehman College, City University of
  New York, NY 10468, USA
}
\author{Michael~G.~Kwakye}
\affiliation{Department of Physics and Astronomy,  Lehman College, City University of
  New York, NY 10468, USA
}

\author{Danijela Kyriakakis}
\affiliation{Department of Physics and Astronomy,  Lehman College, City University of
  New York, NY 10468, USA
}

\author{Krystal Reynoso}

\affiliation{Department of Physics and Astronomy,  Lehman College, City University of
  New York, NY 10468, USA
}

\author{Luis A. Anchordoqui}

\affiliation{Department of Physics and Astronomy,  Lehman College, City University of
  New York, NY 10468, USA
}

\affiliation{Department of Physics,
 Graduate Center, City University
  of New York,  NY 10016, USA
}

\affiliation{Department of Astrophysics,
 American Museum of Natural History, NY
 10024, USA
}

\begin{abstract}
  \vskip 2mm \noindent
Millimeter-waveband spectra of Venus from both the James Clerk Maxwell
Telescope (JCMT)  and the Atacama Large Millimeter/submillimeter Array
(ALMA) seem to indicate there may be evidence (signal-to-noise ratio of
  about $15\sigma$) of a phosphine absorption-line profile against the thermal background from deeper,
hotter layers of the atmosphere. Phosphine is an important biomarker; e.g.,
  the trace of phosphine in
  the Earth's atmosphere is unequivocally associated with anthropogenic
  activity and microbial life (which produces this highly reducing gas
  even in an overall oxidizing environment). Motivated by the JCMT and ALMA
  tantalizing observations we reexamine whether Venus
  could accommodate Earthly life. More concretely, we hypothesize that the 
  microorganisms populating the venusian atmosphere are not free
  floating but confined to the liquid environment inside cloud
  aerosols or droplets. Armed with this hypothesis, we generalize a
  study of airborne germ transmission to
 constrain the maximum size of droplets   that could be floating in the
 venusian atmosphere by demanding that their
 Stokes fallout times to reach moderately high
 temperatures are pronouncedly larger than the microbe's replication
 time.
 We also comment on the effect of cosmic ray
 showers on the evolution of aerial microbial life.
\end{abstract}
\date{April 2024}
\maketitle

\section{Introduction}

It has long been known that the surface of Venus is too harsh an
environment for life~\cite{Sagan:1961,Sagan:1967}. Observations from Mariner 2 and
Venus 4 spacecraft seem to indicate that the surface of Venus is a hot
dielectric carrying temperatures of roughly $427^\circ{\rm C}$. The critical point
of water is $374^\circ{\rm C}$ and 218~atmospheres, and hence liquid water at the
average Venus surface temperature is out of the question. Furthermore,
at temperatures well below $427^\circ{\rm C}$ enzymes are speedily inactivated,
proteins denatured, and most biological organic molecules
pyrolized. All in all, it seems quite safe to conjecture that the mean
surface temperature of Venus excludes terrestrial forms of life.

Contrariwise, it has long been speculated that the clouds of Venus
offer a favorable habitat for life, but regulated to be domiciled at
an essentially fixed altitude~\cite{Morowitz:1967}. The archetype living thing being the
spherical hydrogen gasbag isopyenic organism (sHgio).  The average  
sHgio size can be obtained by demanding 
 that its mass must be equal to the displaced mass of the atmosphere,
 {\it viz.},
 \begin{equation}
   \rho_{\rm H} \ R_2^3 + \rho_{\rm skin} \ (R_1^3-R_2^3) = \rho_{\rm
     atm} \
   R_1^3 \,,
   \label{uno}
\end{equation} 
where $\rho_{\rm atm} \sim 7 \times 10^{-4}~{\rm g/cm}^3$ and $\rho_{\rm H} \sim 5 \times 10^{-5}~{\rm g/cm}^3$ are
respectively the atmospheric and hydrogen densities at
0.5~atm pressure level, and $\rho_{\rm skin} =1.1~{\rm g/cm}^3$ is the
density of the skin modelled as a 
membrane of outer radius $R_1$ and inner radius $R_2$.  For a typical 
skin thickness of roughly $1~\mu{\rm m}$, i.e.,
\begin{equation}
  \frac{R_1-R_2}{R_1} \simeq 2 \times 10^{-4} \,,
\end{equation}
Eq.~(\ref{uno}) implies that sHgios would have an average diameter of $2 R_1 \sim 4~{\rm cm}$.

Millimeter-waveband spectra of Venus from both the James Clerk Maxwell
Telescope (JCMT)  and the Atacama Large Millimeter/submillimeter Array
(ALMA) telescopes show conclusive evidence (signal-to-noise ratio 
  $\sim 15\sigma$) of a phosphine (chemical formula PH$_3$)
absorption-line profile against the thermal background from deeper,
hotter layers of the
atmosphere~\cite{Greaves:1}. Data
reanalyses to address some critiques questioning the bandpass
calibration~\cite{Villanueva}, statistics on flase positives~\cite{Thompson}, and SO$_2$ contamination~\cite{Akins,Lincowski} were presented
in~\cite{Greaves:2,Greaves:3,Greaves:4}.

The PH$_3$
signal has also been claimed to be present in historical data collected by the
Pioneer Venus Large Probe Neutral Mass
Spectrometer~\cite{Mogul}. However, this is a very doubtful
detection. The mass resolution of this instrument together with the
limited data transmitted clearly does not allow to identify PH$_3$ on $m/z$ 34.\footnote{In mass spectrometry, $m/z$ represents the ratio of an ion's mass ($m$) to its charge ($z$).}  This could very well be $^{34}$S or even H$_2$S. Especially $^{34}$S as a fragment of all the sulfur bearing species in the atmosphere of Venus is probably more abundant than PH$_3$. 

The punch line of a potential PH$_3$ detection is that phosphine is a
biosignature gas associated with anaerobic ecosystems~\cite{Sousa-Silva}. Thus, the JCMT and ALMA intriguing observations have reinvigorated investigations looking into the
possibility of life in the atmosphere of
Venus~\cite{Bains,Seager,Clements,Bains:2024}. In particular, it was proposed
in~\cite{Seager} that microbial life could reside inside liquid
droplets/aerosols, which could protect the microbes from a fatal net loss of liquid to the atmosphere, an
unavoidable problem for any free-floating living thing. However, the
aerosol habitat could only have a limited lifetime because it would inexorably
grow into droplets of a large enough
size that are forced by gravity to settle downward to hotter,
uninhabitable layers of the atmosphere. In this paper we generalize a
  study of airborne coronavirus transmission~\cite{Anchordoqui} to
 constrain the maximum size of the droplets that could be floating in the
 venusian atmosphere by demanding that  their
 Stokes fallout times to reach moderately high
 temperatures are pronouncedly larger than the microbe's replication
 time.

 The layout of the paper is as follows. In Sec.~\ref{sec:2} we provide a concise
 summary of the aerial life cycle put forward in~\cite{Seager}.  In
 Sec.~\ref{sec:3} we compare the Stokes fallout time of droplets with
 typical microbe's replication times on Earth (under optimum conditions) to determine the maximum allowed
 radius for the life cycle to persist indefinitely in the atmosphere of
 Venus. In Sec.~\ref{sec:4} we comment on the effect of cosmic ray
 showers on the evolution of aerial microbial life. In
 Sec.~\ref{sec:5} explore whether some kind of organic chemistry could
 develop in the concentrated sulfuric acid of Venus' clouds. The paper wraps up
 in Sec.~\ref{sec:6} with conclusions.

\begin{figure*}[tpb].
  \postscript{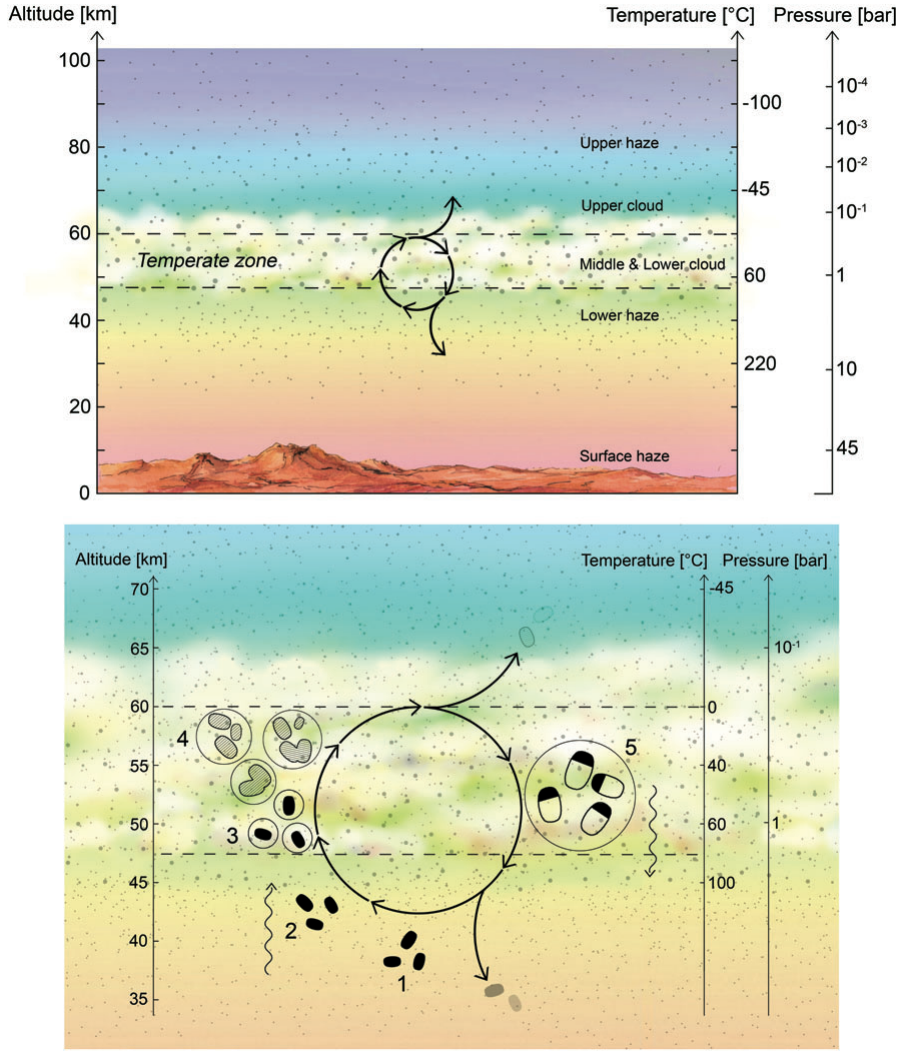}{0.9}
\caption{Hypothetical life cycle of the venusian microorganisms. Top
  panel: Cloud cover on Venus is permanent and continuous, with the
  middle and lower cloud layers at temperatures that are suitable for
  life. Bottom panel: Cycle for venusian aerial microbial life; see
  text for details. Taken from Ref.~\cite{Seager}. \label{fig:1}}
\end{figure*}

\section{Life Cycle for Venusian Aerial Microbes}

\label{sec:2}

Venus' clouds encircle the entire planet, resulting in a high
planetary albedo $\sim 0.8$~\cite{Marov}. The base
of this thick cloud cover is situated at roughly 47~km above the
surface (the temperature at this position is around $100^\circ$C) and extends up to over 70~km in
altitude. In equatorial and mid-latitudes the cloud top is located at
74~km, but decreases towards the poles to roughly 65~km~\cite{Ignatiev}. The size distribution of aerosol particles drifting inside the
clouds enable a subdivision into three layers: upper  (56.5 to 70~km
altitude), middle (50.5 to 56.5~km), and lower (47.5 to 50.5~km); the
smallest type-1 droplets (with a radius of $0.2~\mu{\rm m}$) and type-2 droplets
(with a radius of 1 to 2~$\mu{\rm m}$) are present in all three cloud
layers, whereas the largest type-3 droplets (with a radius of
$4~\mu{\rm m}$) are only present in the middle and lower cloud layers~\cite{Knollenberg}.

It has long been suspected that the cloud decks of Venus offer an aqueous habitat where
microorganisms can grow and flourish~\cite{Grinspoon}. Carbon dioxide, sulfuric acid compounds, and
ultraviolet (UV) light could give microbes food and energy. It is
generally accepted that
existence of life is most likely at an altitude of
50~km, where the temperature is between 60 and 90 degrees Celsius (140 and
194 degrees Fahrenheit), and the pressure is about 1 atm. An
 optimist might even imagine that the microbial life actually arose in a good-natured surface
habitat, perhaps in a primitive ocean, before the
planet suffered a runaway greenhouse, and the microbes lofted into the
clouds~\cite{Schulze-Makuch}. On the other hand, a recent study seems
to indicate that the planet has never been liquid-water habitable~\cite{Constantinou}. 

Even though the Earth's atmosphere does not provide long-lived aerosols with
temperatures or acidities comparable to Venus, the presence of life at
high altitudes has been well-documented:  bacteria, pollen, and algae have been observed in the
Earth's atmosphere as high as 15~km~\cite{Womack}. Furthermore, evidence has been found
for the growth
of bacteria in droplets sampled from a super-cooled cloud near a
meteorological station on a mountain top in the Alps~\cite{Sattler}. It is commonly understood that
 these microorganisms would have likely reached such heights through
 evaporation, storms, eruptions, or meteor impacts. Of course, all of  these
 processes could also have occurred in the early history of Venus. In
 contrast to the Earth, venusian clouds are not transient entities but
 represent a global, uninterrupted phenomenon, with the potential for
 aerosol particles to be sustained for a long period of time, rather
 than just a few days as in the terrestrial atmosphere. Thereby, the venusian
 clouds could provide a stable niche if microbes remain lofted in this aerial. 
 
A cycle for venusian aerial microbial life was developed
in~\cite{Seager}. The general idea, which is shown
in Fig.~\ref{fig:1}, follows five steps that can be
summarized as follows:
\begin{enumerate}[noitemsep,topsep=0pt]
\item The cycle begins with dormant desiccated spores
  (black blobs in Fig.~\ref{fig:1}) which partially populate the lower
  haze layer of the atmosphere.\footnote{In the spirit
    of~\cite{Seager}, throughout we use the term ``spore'' to indicate a cell in a dormant state of long-term metabolic inactivity, which is further resistant to (and protected from) environmental stresses.} 
\item Updraft of spores transports them up to the habitable
  layer. For example, the spores could travel up to the clouds via the
  effect of gravity waves. Despite the fact that gravity waves can
  only lead to the net transport of energy and momentum and not
  matter, they can compress atmosphere layers as they travel,
  producing vertical winds (which have been measured directly by the
  Venera landing probes 9 and 10 at the atmospheric lower haze layers~\cite{Kerzhanovich}).
\item  Shortly after reaching the (middle and lower cloud) habitable layer, the spores
  act as cloud condensation nuclei, and once surrounded by liquid (with necessary chemicals dissolved) germinate
  and become metabolically active.
\item  Metabolically active
  microbes (dashed blobs in Fig.~\ref{fig:1}) grow and divide within liquid droplets
  (shown as solid circles in the figure). The liquid droplets grow by coagulation.
\item The droplets reach a size large enough to gravitationally
  settle down out of the atmosphere; higher temperatures and droplet
  evaporation trigger cell division and sporulation. The spores are
  small enough to withstand further downward sedimentation, remaining
  suspended in the lower haze layer
(a depot of hibernating microbial life) to restart the cycle.
\end{enumerate}

One of the key assumptions of the aerial life cycle put forward in~\cite{Seager} is the timescale on
which droplets would persist in the habitable layer to empower
replication. It is this that we now turn to study.

\section{Replication rates and Fallout times}

\label{sec:3}

Self-replication is a faculty inherent to every species of living
thing, and standard intuition imposes that such a physical process
must invariably be fueled by the production of
entropy~\cite{Chudnovsky:1985}. It is of interest then to estimate a
lower bound for the amount of heat that is produced during a process
of self-replication in a system coupled to a thermal bath.  The
minimum value for the physically allowed rate of heat production is
determined by the growth rate, internal entropy, and durability of the
replicator. Impressively, bacteria replicate close (within a factor of
two or three) to the physical efficiency~\cite{England:2012}.

 Bacteria replicate by binary fission, a process by which one
 bacterium splits into two. Therefore, bacteria increase their numbers
 by geometric progression whereby their population doubles every
 generation time. Generation time is the time it takes for a
 population of bacteria to double in number. For many common bacteria,
 the generation time is quite short, 20 to 60 minutes under optimum
 conditions. The typical example is Escherichia coli (or E. coli),
 which can divide every 20 minutes under aerobic, nutrient-rich
 conditions (but of course the generation time for bacteria in the
 wild are substantially greater than those in the laboratory)~\cite{Gibson}.
 Actually, Vibrio natriegens (previously known as Pseudomonas
 natriegens and Beneckea natriegens) is a free-living marine bacterium
 with the fastest generation time known, between 7 to 10
 minutes~\cite{Eagon,Maida,Lee}. In our calculations we adopt as
 benchmark the E. coli generation time
 under optimum conditions. The duplication times of other types of
 bacterium are listed in Table~\ref{table}.
 
 \begin{table}
   \caption{Bacteria generation times. \label{table}} 
\begin{tabular}{lcr}
  \hline
  \hline
  Name & Reproduction Time & Ref. \\
\hline
         Bacillus subtilis &
20 minutes & \cite{Errington}\\
Pseudomonas aeruginosa & 
16 to 24 hours & \cite{LaBauve} \\
Vibrio cholerae &
                  20 minutes & \cite{Das}\\
  Bacillus thuringiensis &
                           20 minutes & \cite{Choi} \\
  Shigella flexneri &
                      40 minutes & \cite{Jennison} \\
  Streptococcus pyogenes &
12 to 16 hours & \cite{Gera} \\
Salmonella typhimurium &
20 minutes & \cite{Lowrie} \\
Clostridium perfringens &
~~~~~~~~~10 to 12 minutes~~~~~~~~~ & \cite{Li} \\
Pseudomonas fluorescens &
                          1.5 hours & \cite{Caldwell} \\
  \hline
  \hline
\end{tabular}
\end{table}

The mean division time for bacteria population is then 20 minutes. If the
observation begins with one bacterium, we can estimate how many
bacteria will be present after six hours. The E. coli divides every 20
minutes, and so this bacterium divide (60/20 =3) three times every
hour. If the bacteria grow for twelve hours, each bacterium will divide 3
times per hour $\times$ 12 hours = 36 times. Every time the bacteria
reproduce, the number doubles. Then, the number of bacteria at the end
of the growth period is found to be
\begin{equation}
N_{\rm final} = N_{\rm initial}  \ 2^n   \,,
\end{equation} 
where $N_{\rm initial}$ is number of bacteria at the beginning of the
growth period and  $n$ is the number of divisions. For $N_{\rm initial} = 1$ and $n =36$, we have
$N_{\rm final} \sim 7 \times 10^{10}$ bacteria. We conclude that to maintain
the colony of microbes alive we require Stokes fallout
times longer than half an Earth day.

Assuming that the droplets are spherical  the mass can be simply estimated as
\begin{equation}
m_{\rm droplet} = \frac{4}{3} \pi r_{\rm droplet}^3 \ \rho \,,
\end{equation}
where $\rho$ is the droplet's density; in our calculations we consider the  E. coli dry mass density $\rho_{\rm Ec} \sim
300~{\rm kg/m}^3$~\cite{Pang} and the density of water $\rho_{\rm H_2O} \sim
997~{\rm kg/m}^3$.

Under the action of gravity, droplets of mass $m_{\rm droplet}$ and size
$r_{\rm droplet}$ would acquire a downward terminal speed that follows from Stokes law and is given by
\begin{equation}
v_{\rm down} = \mu \ m_{\rm droplet} \ g_{_{\venus}} \,,
\label{vdown}
\end{equation}
where $g_{_{\venus}}$ is the acceleration due to gravity in Venus,
\begin{equation}
  \mu = \frac{1}{6\pi \ \eta_{_{\venus}} \ r_{\rm droplet}}
  \label{mu}
\end{equation}  
is the organism mobility in the fluid, and where $\eta_{_{\venus}}$ is
the dynamic viscosity of the venusian atmosphere~\cite{Einarsson}. For
the dynamic viscosity, we adopt the  estimate of Ref.~\cite{Beck},
$\eta_{_{\venus}} \sim 2 \times 10^{-5}~{\rm kg/m/s}$.

\begin{figure}[t]
  \postscript{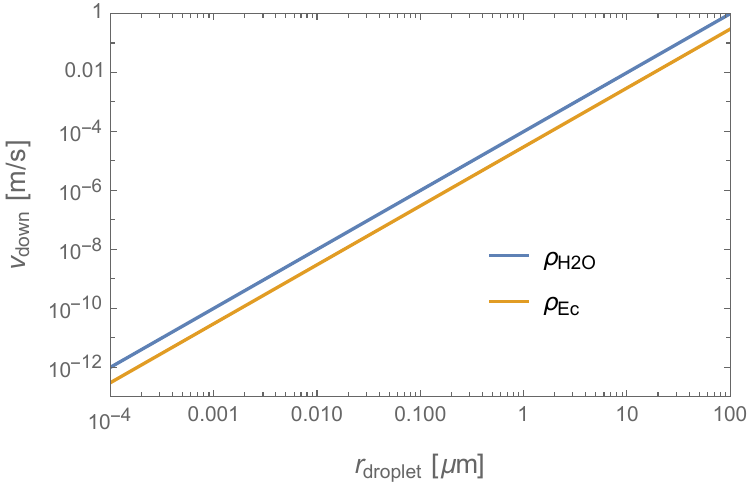}{0.9}
\caption{Terminal speed as a function of the microbe's radius. \label{fig:2}}
\end{figure}

The acceleration of gravity at the surface of Venus can be derived
from measurements of Mariner V~\cite{Anderson:1967}
\begin{equation}
  G_N M_{_{\venus}} = 324,856.6 \pm 0.5~{\rm km^3/s^2} \,,
\end{equation}
and is found to be $g_{_{\venus}} \simeq 8.9~{\rm m/s^2}$, where $G_N$ is
Newton's gravitational constant and where we have taken the
radius of Venus to be $R_{_{\venus}} \simeq 6,056~{\rm km}$ to
accommodate existing 
observations~\cite{Ash:1967,Ash:1968,Anderson:1968}.

Putting all this together, we can now combine Eqs.~(\ref{vdown}) and (\ref{mu}) to estimate the downward
terminal speed of the droplets. In Fig.~\ref{fig:2} we show the terminal
velocity as a function of the droplet's radius. As one
can check by inspection, the terminal speed of aerosols ({\it viz.} droplets
with $r_{\rm droplet} \alt 1~\mu{\rm m}$) is negligible, and so we
can conclude that gravity would play no role in the motion of the
microbes through the atmosphere. As the droplet size approaches
$100~\mu{\rm m}$  the droplets would start sinking to the lower haze
layers. For example, a droplet with density $\sim \rho_{\rm Ec}$ and a
radius $\sim 100~\mu{\rm m}$ would attain a terminal speed $v_{\rm
  down} \sim 0.1~{\rm m/s}$, and thus would fall about 4~km in half an
Earth day.

 \section{Cosmic ray effects on microbial life}
\label{sec:4}

We have long been suspecting that cosmic radiation played a
pivotal role
in the evolution of life on Earth~\cite{Joly}; for a more recent
perspective see~\cite{Dartnell}. In this section, we briefly comment on
whether energetic particle radiation could provide a threat for any
type of
microbe populating the clouds of Venus.

Pioneer Venus measurements are consistent with the absence of an intrinsic magnetic $\vec B$ field that could provide a
 shield against low and
 high energy charged particles reaching the atmosphere~\cite{Phyllips,Russell}. If taken at
 face value, the lack of $\vec B$ field seems to indicate that
 the solar wind would interact directly with the upper
 atmosphere. However, the upper atmosphere is ionized by UV radiation,
 yielding the so-called ``ionosphere.'' Currents arising from the interaction
 between the solar wind and the electrically conductive venusian
 ionosphere induce a magnetic field, and so the incoming solar wind
 particles are slowed down and diverted around the
 planet~\cite{Zhang}. However, the induced magnetic field is
 weak~\cite{Russell:2006} and can
 only deflect charged energetic particles with energies up to several
 hundreds of keV. Thus, most of the cosmic radiation have
 unrestricted access to the Venusian
 atmosphere.

Cosmic rays with energies above about
1~GeV induce particle cascades via inelastic scattering with
atmospheric nuclei, producing fluxes of secondary, tertiary, and
subsequent generations of particles. All these particles together
create a cascade (or shower). As the cascade develops longitudinally the particles become less and less energetic since the energy of the incoming cosmic ray is redistributed among more and more participants. The transverse momenta acquired by the secondaries cause the particles to spread laterally as they propagate through the atmospheric target~\cite{Anchordoqui:2018qom}.
 
As can be seen in Fig.~\ref{fig:1}, the top region of the temperate
zone is at 62~km altitude, and thus it receives less than $200~{\rm g/cm^2}$
shielding depth against cosmic ray showers. This is far less than the
biosphere on the surface of the Earth, which is beneath $1033~{\rm
  g/cm^2}$. The atmospheric grammage at the middle cloud layer is, however,
$1000~{\rm g/cm^2}$, suggesting the cosmic ray induced effects would
be comparable to those on the
Earth's surface. Indeed, numerical simulations show that cosmic radiation would
not have had any hazardous effect on putative microorganisms within the
potentially temperate zone (51 to $62~{\rm km}$)~\cite{Dartnell:2015,Herbst,Sciutto:2025ilz}.

\section{Life Outside the Habitable Zone}
\label{sec:5}

A habitable planet, in the context of exoplanets, is a planet that has
the potential to support life, typically defined by its location
within a star's habitable zone, where temperatures are suitable for
liquid water (a key ingredient for life as we know it) to exist on the
surface. These are the so-called ``Goldilocks planets.'' A more
a broader and certainly nontypical definition of a habitable planet is a frontier to be
explored and requires pushing the boundaries of our terracentric
viewpoint for what we deem to be a habitable
environment~\cite{Lingman,Anchordoqui:2019et,Hoehler,Anchordoqui:2020wsb}. In this section, we relax our assumption
about Earthly life to briefly discuss recent findings which suggest life may be supported in the extreme
solvent conditions of the Venusian acid
clouds~\cite{Bains:2021a,Bains:2021b}.

The study of sulfuric acid hydrolysis of proteins dates back to
1820~\cite{Braconnot}. Subsequently, studies continued to explore the
reactivity of biological matter in concentrated sulfuric acid (and
other acids). This pioneer work searched for the chemical
composition of biological material and eventually attempted to
decipher the amino acid sequence in protein polymers. Such studies on
the reactivity of proteins in concentrated sulfuric acid aimed to
chemically split peptide bonds at specific amino acid positions in
polypeptides~\cite{Reitz,Ramachandran}. Today we know that a rich organic chemistry can evolve from simple precursor molecules seeded into concentrated sulfuric acid.

The sulfuric acid concentration in the clouds of Venus has not
been directly measured, but instead inferred from Pioneer Venus
measurements of gases by the mass spectrometer~\cite{Hoffman}. The gases evolved from cloud
particles that clogged the inlet and are consistent with cloud
droplets composed of 15\% w/w H$_2$O and 85\% w/w H$_2$SO$_4$.\footnote{The weight concentration of a solution is expressed
  as w/w, which stands for weight-per-weight. The volume of each
  chemical is disregarded and only the weight counts.} Based on models, it is likely that the sulfuric acid
concentration of the cloud particles varies with altitude, in the cloud tops reaching 79\%
w/w while in the lower clouds the sulfuric acid concentration could reach 98\% w/w~\cite{Krasnopolsky}.

It has been shown that the majority of amino acids are stable in the
range of Venus' cloud sulfuric acid concentrations (81\% and 98\% w/w, the rest being
water)~\cite{Seager:PNAS,Seager:M,Seager:2024life}. Another step towards true biochemistry in this
aggressive solvent has been the observation of stability of lipid
membranes~\cite{Duzdevich}. Finally, some dipeptides, which are precursors to larger peptides and proteins,
could also be stable at both sulfuric acid concentrations for many
months, if not longer~\cite{Petkowski:2024,Petkowski:2025}. The stability of nucleic
acid bases and lipids in concentrated sulfuric acid advances the idea
that chemistry to support life may exist in the Venus cloud particle
environment.

In closing, we note that our estimate of the Stokes fall-out times
required for droplets of sulfuric acid to reach the lower haze layers of the
atmosphere remains well grounded, because the density of sulfuric
acid, $\rho_{\rm H_2SO_4} \sim 1830~{\rm kg/m^3}$, is only a factor of
two larger than that of water.
However, whether organic chemistry in concentrated sulfuric acid would
allow the bacterial replication
time to be shorter than the microbe's lifetime remains an open question.

 \section{Conclusions}
\label{sec:6}

The potential detection of of PH$_3$ absorption lines in Venus' 
atmosphere, a gas often associated with biological processes on Earth,
has sparked excitement and renewed interest in the possibility of life
beyond Earth. At present, the scientific community is divided into those who support the detection of phosphine~\cite{Greaves:1,Greaves:2,Greaves:3,Greaves:4} and those
who question the measurements and the presence of
PH$_3$~\cite{Villanueva,Thompson,Akins,Lincowski}. Certainly,
more data are needed to shed light on this conundrum.

From the perspective of terrestrial chemistry, phosphine is considered
a biogenic molecule, typically produced by living organisms. Possible
pathways for PH$_3$ production in a Venusian environment were
investigated, with the conclusion that the observed abundance of PH$_3$ cannot
be explained by conventional gas-phase reactions, surface and
subsurface geochemical reactions,
photochemistry, and other non-equilibrium
processes~\cite{Bains,Bains:2024}. The observed PH$_3$ absorption
lines must then originate in some unknown geochemistry/photochemistry
process. An even more extreme possibility is that a strictly aerial
microbial biosphere is responsible for the observed PH$_3$ signal.

Even though the detection of a PH$_3$ signal is yet under debate, it
is interesting to entertain the possibility that it could be the
footprint of aerial microbial life. A proposed Venusian life cycle
suggests that microbial life could exist within the lower cloud
layers, cycling between a dormant, spore-like state in the haze and an
active, metabolically active state within cloud droplets~\cite{Seager}. We have
re-examined whether one can actually envisage life inside the clouds
of Venus, operating entirely on known terrestrial principles. We have
shown that for aerosols, the Stokes fallout times to reach the lower haze
atmospheric layers is pronouncedly larger than the typical bacterium replication time on
Earth. Bearing this in mind, we can conclude that if updraughts exist, a stable population
of microorganisms that in the early history of Venus emigrated from
the surface to the atmospheric clouds and now remain confined
to aerosols may be possible. However, as noted in~\cite{Constantinou} caution must be
taken. If Venus
was once habitable we could forecast a water-rich interior of the
planet, which can be probed inspecting its volcano activity. Now, the gases that Venus' volcanoes are releasing
are about 6\% water, whereas expectation from Earth-like volcanism in
Venus would lead to about 96\% of water.
Thus, the dryness we observe today in the interior of Venus seems to be
inconsistent with the planet ever having oceans. Leaving aside our
working hypothesis, we can
argue that if Venus has
always been a hellish hot planet, then organic chemistry developing in
concentrated sulfuric acid may be an alternative. Whether it is Earth-like or not, an aerial microbial life may be awaiting  for the Venus Life Finder
(VLF) Missions to arrive~\cite{Seager:2021m}.

\section*{Acknowledgments}

We thank William Bains and Janusz Petkowski for permission to
reproduce Fig.~\ref{fig:1}. The research of LAA is supported by
the U.S. NSF grant PHY-2412679.

\end{document}